\documentclass[aps,a4,twocolumn]{revtex4}
\pdfoutput=1
\usepackage{amsmath}
\usepackage{latexsym}
\usepackage{float}
\usepackage{amssymb}
\usepackage{graphicx}
\usepackage{epsfig}

\begin{document}
\title{A fast Monte Carlo algorithm for studying bottle-brush polymers}

\author{Hsiao-Ping Hsu$^1$ and Wolfgang Paul$^2$}

\affiliation{$^1$Institut f\"ur Physik, Johannes Gutenberg-Universit\"at 
Mainz, Staudinger Weg 7, 55099 Mainz, Germany\\
$^2$Theoretische Physik, Martin Luther Universit\"at 
Halle-Wittenberg, von Seckendorff platz 1, 06120 Halle, Germany}

\begin{abstract}
   Obtaining reliable estimates of the statistical properties
of complex macromolecules by computer simulation is a task that
requires high computational effort as well as the development
of highly efficient simulation algorithms.
We present here an algorithm combining local moves, the pivot
algorithm, and an adjustable simulation lattice box for
simulating dilute systems of 
bottle-brush polymers with a flexible backbone
and flexible side chains under good solvent conditions.
Applying this algorithm to the bond fluctuation model,
very precise estimates of the mean square end-to-end
distances and gyration radii of the backbone and side chains
are obtained, and 
the conformational properties of such a complex macromolecule
are studied. 
Varying the backbone length (from $N_b=67$ to $N_b=1027$),
side chain length (from $N=0$ to $N=24$ or $48$), 
the scaling predictions for the backbone behavior as well as 
the side chain behavior are checked. 
We are also able to give a direct comparison of the structure 
factor between experimental data and the simulation results.
\end{abstract}

\maketitle





\section{Introduction}
\label{}

The so-called ``bottle-brush polymers" consist of
a long macromolecule serving as a ``backbone"
on which many flexible side chains are densely 
grafted~\cite{Zhang2005, Subbotin2007, Sheiko2008, Potemkin2009}.
In nature, bottle-brush like aggrecans have been found
in the cartilage of mammalian including human joints
and are indeed held responsible for
the excellent lubrication properties in such 
joints~\cite{Klein2006, Klein2009}.
Recently, the chemical synthesis of such complex molecular
bottle-brushes has become possible in laboratories with 
newly developed synthetic 
techniques~\cite{Wintermantel1994, Wintermantel1996, Rathgeber2005,
Zhang2006}.
Theoretical predictions [11-42]
of the conformational properties of 
bottle-brush polymers based on the blob picture, 
the scaling theory, and the self-consistent 
field theory have also been worked out. 
However, in order to check the theoretical predictions and to give 
a reasonable explanation for the experimental results or
to control the functions of bottle-brush polymers 
computer simulations are needed for a deeper understanding of
the structure of these macromolecules.


{``Static" Monte Carlo (MC) algorithms 
(simple sampling of self-avoiding walks (SAWs)),
extensions such as dimerization, enrichment techniques, Rosenbluth's
inversely restricted sampling and the pruned enriched Rosenbluth
method (PERM) do not converge for very large branched polymers such
as bottle-brushes. ``Dynamic" MC algorithms also
encounter serious problems, since the relaxation times of these polymers
are expected to be excessively large, and hence prohibitively long MC 
simulations would be required.}
In the early work on bottle-brush polymers in
a good solvent, both lattice and off-lattice models were used
for MC simulations. 
Using the bond fluctuation model
~\cite{Carmesin1988, Deutsch1991, Paul1991, Binder1995}
on a simple cubic lattice, applying local moves to
the bottle-brush polymers, side chain lengths up to $N=64$, 
backbone lengths up to 
$N_b=64$, and grafting densities up to $\sigma=1$ were studied 
in~\cite{Shiokawa1999}. In combination with the 
pivot move algorithm, $N_b$ up to $800$, $N$ up to $80$,
and $\sigma < 1/3$ were studied in~\cite{Rouault1996, Rouault1998}.
Using flexible freely jointed chains~\cite{Saariaho1997, Yethiraj2006} 
and the bead-spring model~\cite{Elli2004} with hard sphere interactions
in a continuous space
and using the pivot moves, the largest 
bottle-brush polymers which were studied were $N_b=402$, $N=25$, and 
$\sigma=1$, and using the Metropolis algorithm for the latter model,
maximum values of $N_b=100$, $N=50$ were studied.
{However, although all these studies clearly are very interesting
and stimulating, the accessible parameter range clearly was not large
enough (and the accuracy of the results not precise enough) to allow
a straightforward test of the theoretical concepts, and thus a need
for further work with different methods clearly did emerge.}

In our previous work~\cite{Hsu2007a, Hsu2007b}, we were able to
simulate bottle-brush polymers with rigid linear backbone and flexible
side chains under various solvent conditions
by applying a variant of the PERM algorithm~\cite{Grassberger1997} to 
a simple lattice model with periodic boundary conditions
along the backbone.
Under good solvent conditions, we have shown that the power laws 
predicted for the side chain behavior are still difficult to reach 
although the maximum side chain length in our simulation was $N=2000$.
However, a crossover behavior from  3D SAW-like side
chains to stretched side chains was presented in~\cite{Hsu2007a, Hsu2007b} as 
the side chain length or the grafting density increased.
{Of course, removing all configurational degrees of freedom of
the backbone was a crucial prerequisite to allow the successful use
of the PERM algorithm, but it also means that highly
interesting questions (such as backbone stiffening 
due to the side chains) could not be studied.}

Recently, we have focused on the comparison of structure factors
to experimental data and on the persistence
length~\cite{Hsu2009, Hsu2010a, Hsu2010b} of the backbone using 
the bond fluctuation model
on a simple cubic lattice. This communication will discuss in detail the
algorithm we developed for these simulations. It is introduced in Sec. II, 
and results for the conformational properties of bottle-brush polymers only
achievable by this very efficient simulation approach 
are shown in Sec. III. Finally, our conclusions are presented in Sec. IV.

\begin{figure}[htb]
\begin{center}
\vspace{1cm}
\includegraphics[scale=0.22,angle=0]{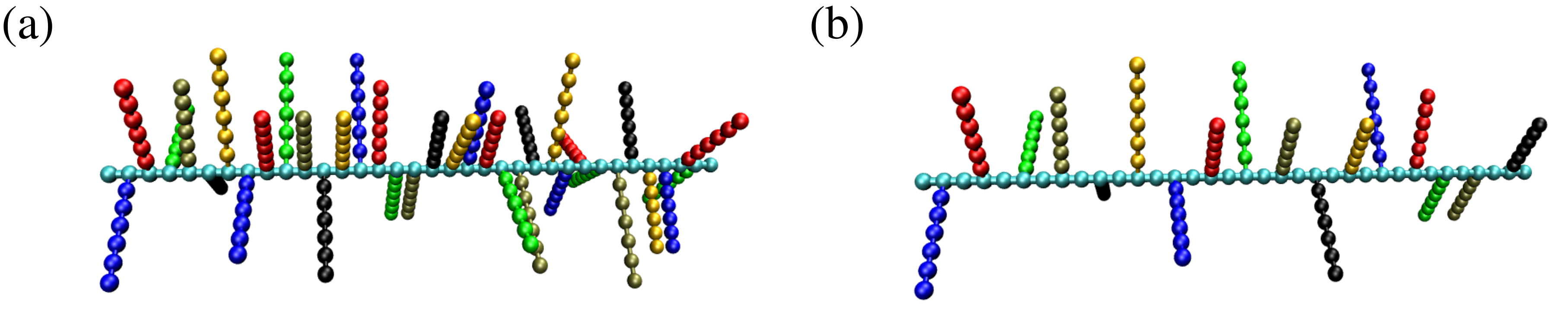}
\caption{Snapshots of initial configurations of bottle-brush polymers
with $N_b=35$ monomers on the backbone and $N=6$ monomers
on each side chain, showing the geometry of our model for
bottle-brush polymers with the grafting densities
(a) $\sigma=1$ and (b) $\sigma=1/2$.}
\label{fig-ini}
\end{center}
\end{figure}

\section{Model and Simulation methods}

For studying bottle-brush polymers with a flexible backbone
and with flexible side chains under very good solvent conditions 
so that only excluded volume effects are considered, 
we generalize the standard bond fluctuation model for linear 
polymers~\cite{Carmesin1988, Deutsch1991, Paul1991, Binder1995} 
to bottle-brush polymers.
In the standard bond fluctuation model, a flexible polymer
chain with excluded volume interactions is described by a chain of
effective monomers on a simple cubic lattice 
(the lattice spacing is the unit of length).
Each effective monomer blocks all 8 corners of an elementary cube
of the lattice from further occupation. Two successive monomers along a
chain are connected by a bond vector $\vec{b}$ which
is taken from the set $\{(\pm 2,0,0)$, $(\pm 2, \pm
1,0)$, $(\pm 2, \pm 1, \pm 1)$, $(\pm 2, \pm 2, \pm 1)$, $
(\pm 3,0,0)$, $(\pm 3, \pm1,0)\}$,
including also all permutations. 
The bond length $| \vec{b} | = \ell_b$ is in a range between 
$2$ and $\sqrt{10}$. There are in total 108 bond vectors serving as
candidates for building the conformational structure
of bottle-brush polymers.

\begin{figure}[htb]
\begin{center}
(a)\includegraphics[scale=0.25,angle=270]{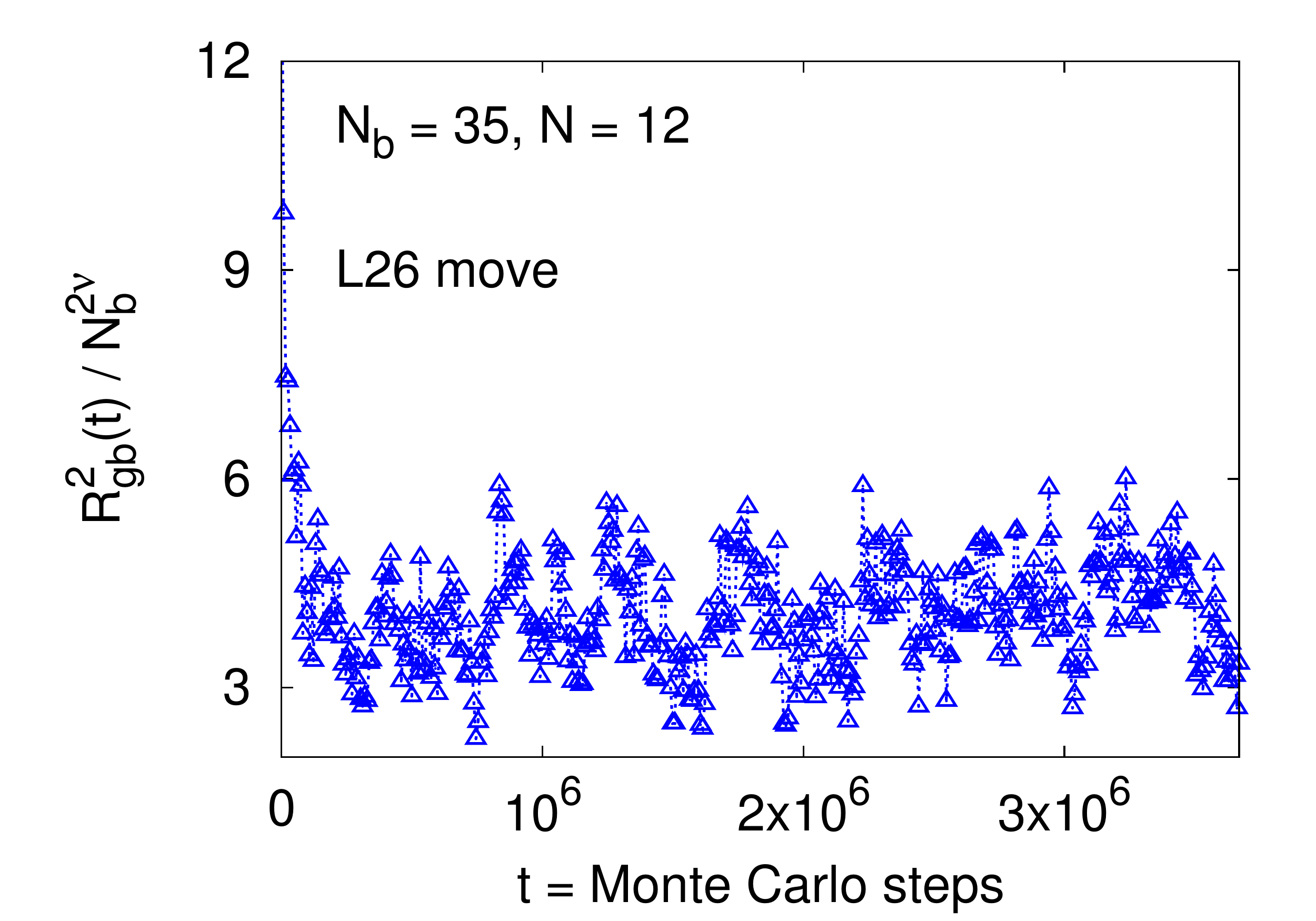}\hspace{0.6cm}
(b)\includegraphics[scale=0.25,angle=270]{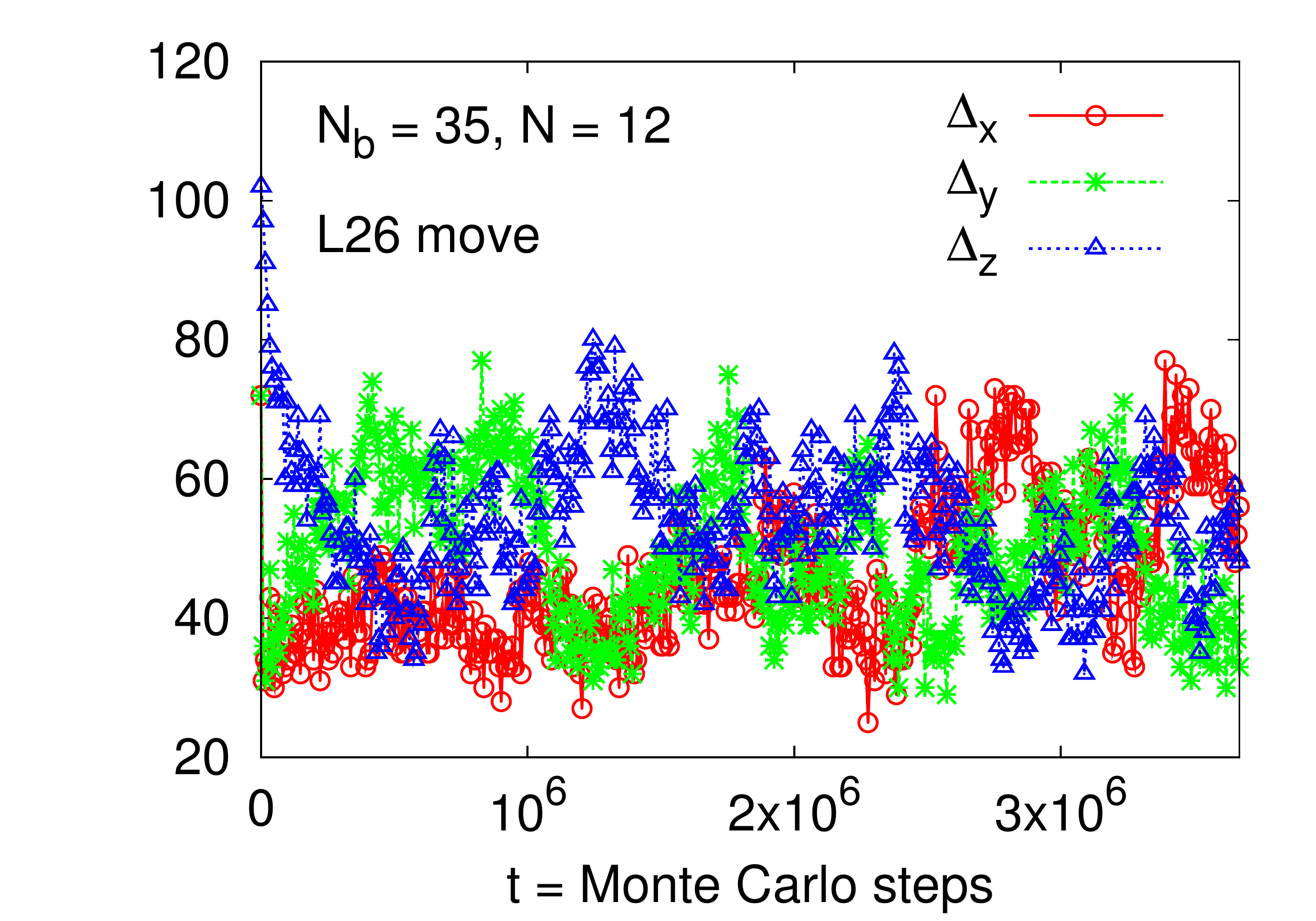}
\caption{\label{fig-bb35-ini}Time series of the rescaled square gyration radii
for the backbone monomers, $R_{gb}^2(t)/N^{2\nu}$ (a),
and the space occupations in the
Cartesian coordinates $(\Delta_x(t), \Delta_y(t), \Delta_z(t))$ (b).
Here we apply the ``L26" move algorithm to simulate the bottle-brush
polymers of $N_b=35$, $N=12$, and $\sigma=1$.}
\end{center}
\end{figure}

As shown in Fig.~\ref{fig-ini}, the geometry of the bottle-brush
polymer is arranged in a way that side chains of length $N$
are added to the backbone 
at a regular spacing $1/\sigma$ ($\sigma$ is the grafting density),
and two additional monomers are
added to the two ends of the backbone.
Thus, the number of monomers of the backbone $N_b$ is
related to the number of side chains $n_c$ via
\begin{equation} \label{eq15}
N_b=[(n_c-1)/\sigma +1 ] + 2 \quad,
\end{equation}
and the total number of monomers of the bottle-brush polymer is
$N_{\rm tot}=N_b+n_c N$.
{Creating an initial configuration of a bottle-brush polymer
that does not violate any excluded volume constraint would be
a highly nontrivial matter if we would require that both backbone
and side chains are already coiled.
Thus} one simple way to construct an initial configuration
of bottle-brush polymers in the simulation is to assume that the backbone
and all side chains are rigid rod-like structures.
In order to fit the criteria of the bond fluctuation model without
further checking, the backbone is placed in the direction along
the $z$-axis with fixed bond length $\ell_b=3$ between two
successive monomers on the backbone.
The bond vectors of each side chain are chosen randomly from one of the
allowed bond vectors in the ($xy$)-plane, but the bond vectors
used to connect the monomers on the same side chain starting
from the grafting site on the backbone are the same.

In our algorithm, instead of trying to move a monomer to the
nearest neighbor sites in the six directions (``L6" move),
$\pm \hat{x}$, $\pm \hat{y}$, and $\pm \hat{z}$ for the 
standard bond fluctuation model, 
we use the local 26 (``L26") moves~\cite{Wittmer2007}. Namely,
the chosen monomer is allowed to move to not only the 
nearest neighbor sites but also the next neighbor
sites, and the sites at the 8 corners, which are located
$\sqrt{2}$, and $\sqrt{3}$ lattice spacings away from the 
chosen monomer, respectively. 
The move is accepted only if the selected new positions
are empty and the bond length constraints are satisfied.
The key point of this move is
that it allows for crossings of bonds during the move; no such
simple moves that allow bond crossing for the simple 
SAW model (where
the bond length is fixed to one lattice spacing) are known.
Without the possibility of bond
crossing, two side chains of the bottle-brush that happen to be
entangled with each other would relax this topological constraint
only extremely slowly.

\begin{figure}[htb]
\begin{center}
(a)\includegraphics[scale=0.25,angle=270]{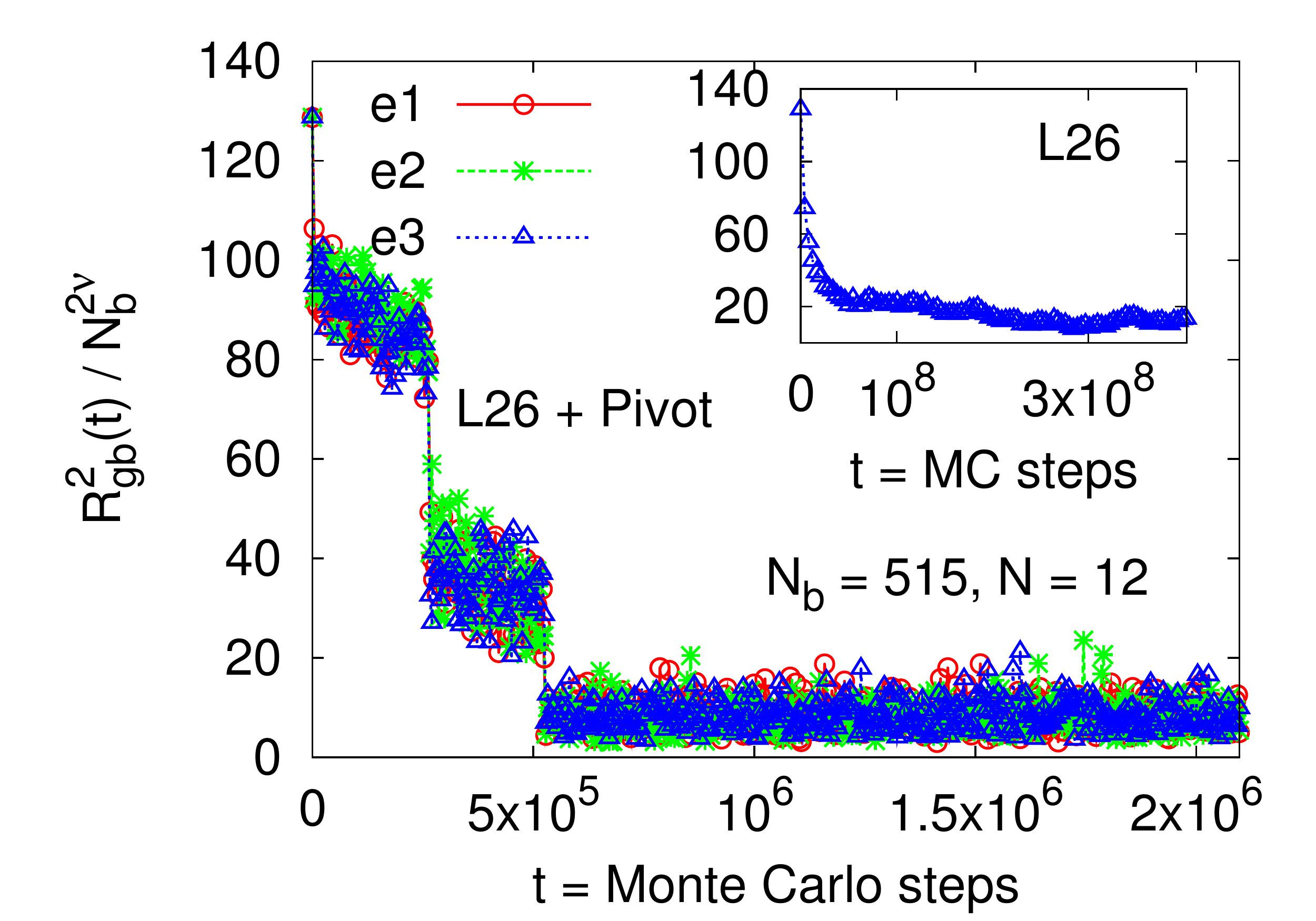}\hspace{0.6cm}
(b)\includegraphics[scale=0.25,angle=270]{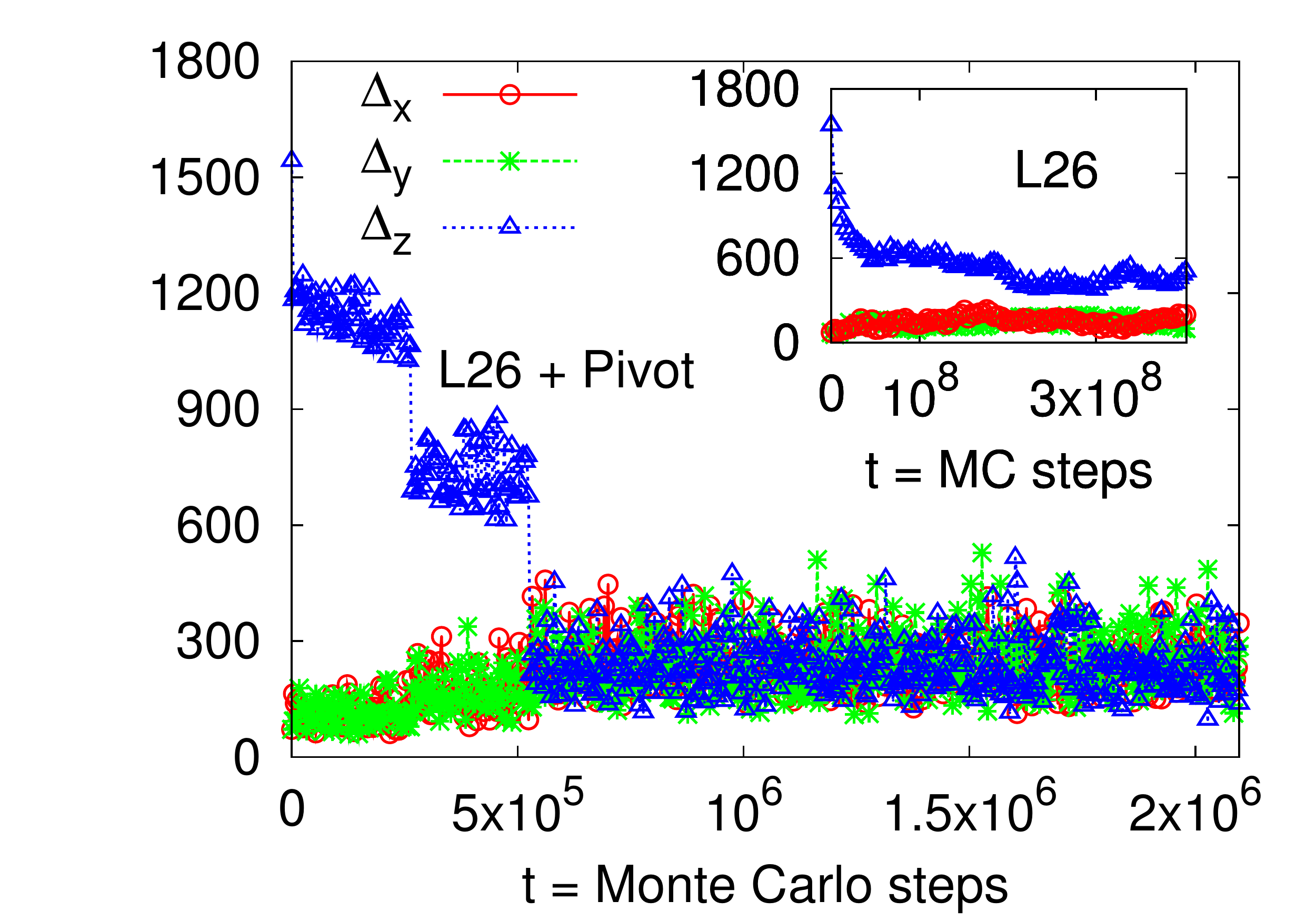}
\caption{\label{fig-bb515-ini}
Time series of the rescaled square gyration radii
for the backbone monomers, $R^2_{gb}(t)/N_b^{2 \nu}$ (a),
and the space occupations in the
Cartesian coordinates $(\Delta_x(t), \Delta_y(t), \Delta_z(t))$ (b).
Here we apply the ``L26 +pivot" algorithm to simulate a
bottle-brush polymer of $N_b=515$, $N=12$, and $\sigma=1$.
The time series obtained when using the ``L26" move is shown in the
inset. The legends $e_1$, $e_2$, $e_3$ in (a)
indicate three different starting configurations.}
\end{center}
\end{figure}

Since neighboring side chains are rigidly 
grafted to neighboring
``anchor points" at the backbone, they can never diffuse away from
each other, unlike free chains in a solution or melt;
thus topological constraints could relax only via an 
``arm retraction mechanism"
familiar from star polymers, if the algorithm ensures 
non-crossability of chains. This arm reaction 
leads to an extremely long relaxation time.
This problem is avoided by the ``L26" moves.
Applying the ``L26" move algorithm to a small bottle-brush polymer
of 431 monomers ($N_b=35$, $N=12$, and $\sigma=1$), results of the
time series of the square gyration radius of the backbone scaled
with the scaling law for 3D SAWs, $R_{gb}^2(t)/N^{2 \nu}$
($\nu=0.588$), and the time series of the shape change of
bottle-brush polymers, which is described by the space occupation
in the Cartesian coordinates $(\Delta_x(t), \Delta_y(t), \Delta_z(t))$
shown in Fig.~\ref{fig-bb35-ini} indicate that it needs about
$10^6$ MC steps to reach an equilibrium state.
Here one MC step is a sequence
of $N_{\rm tot}$ ``L26" moves (each monomer is attempted
to move once).
As the number of monomers on the backbone increases to $N_b=515$,
i.e., the total number of monomers increases to $N_{\rm tot}=6671$,
one would expect for a Rouse scaling of the relaxation time
$\tau \sim N^{1+2\nu}$ that it might need about $400$ times 
the number of MC steps to reach an 
equilibrium state. Unfortunately, the simple Rouse behavior
does not describe the scaling of the relaxation time for the
bottle-brushes and the relaxation time increases even faster
as shown in the inset of Fig.~\ref{fig-bb515-ini}. 


\begin{figure}[htb]
\begin{center}
\vspace{1cm}
\includegraphics[scale=0.25,angle=0]{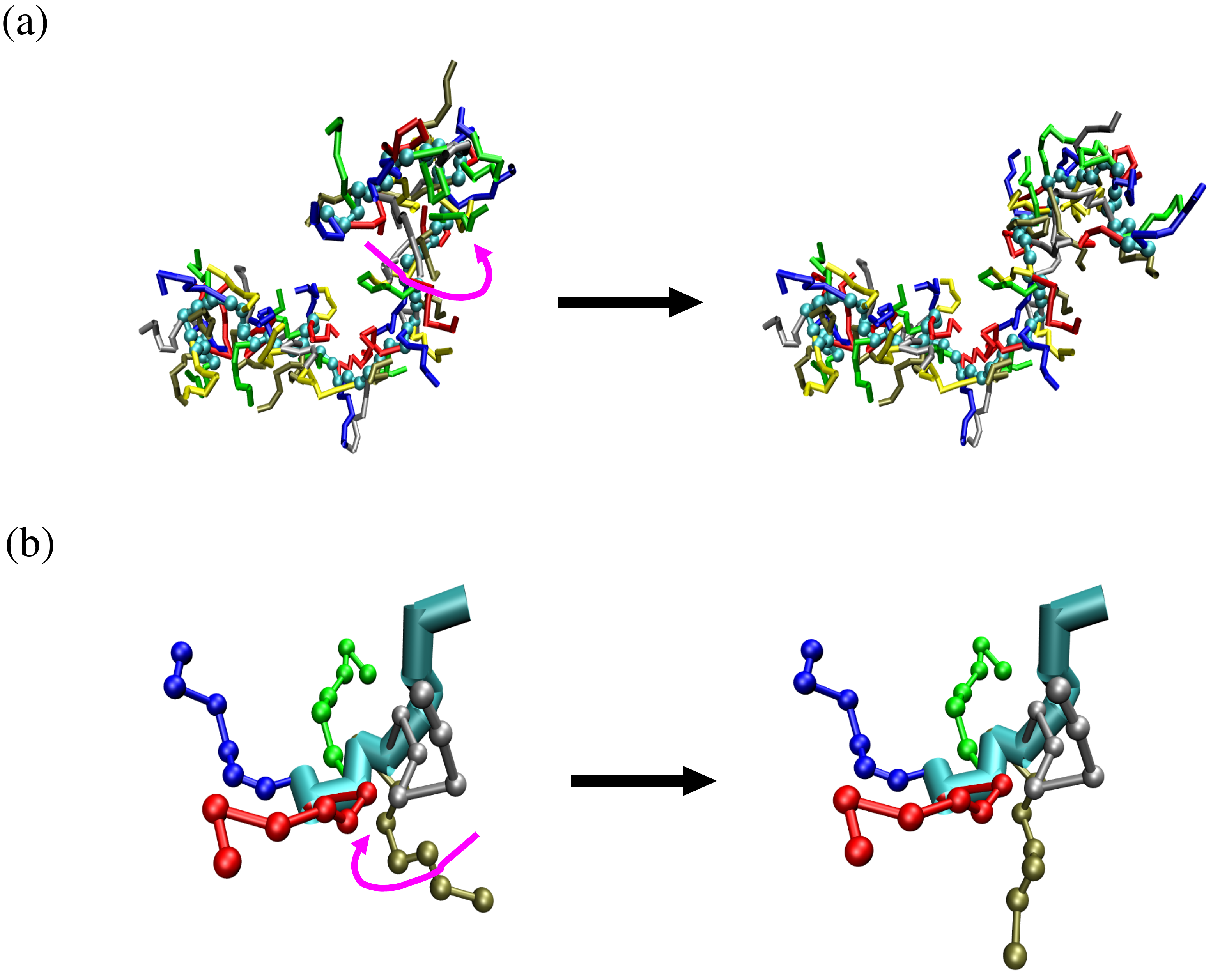}
\caption{Pivot moves applied to a randomly chosen monomer
on the backbone (a) and on one of the side chains (b).}
\label{fig-pivot}
\end{center}
\end{figure}

\begin{figure}[htb]
\begin{center}
\vspace{1cm}
(a)\includegraphics[scale=0.25,angle=270]{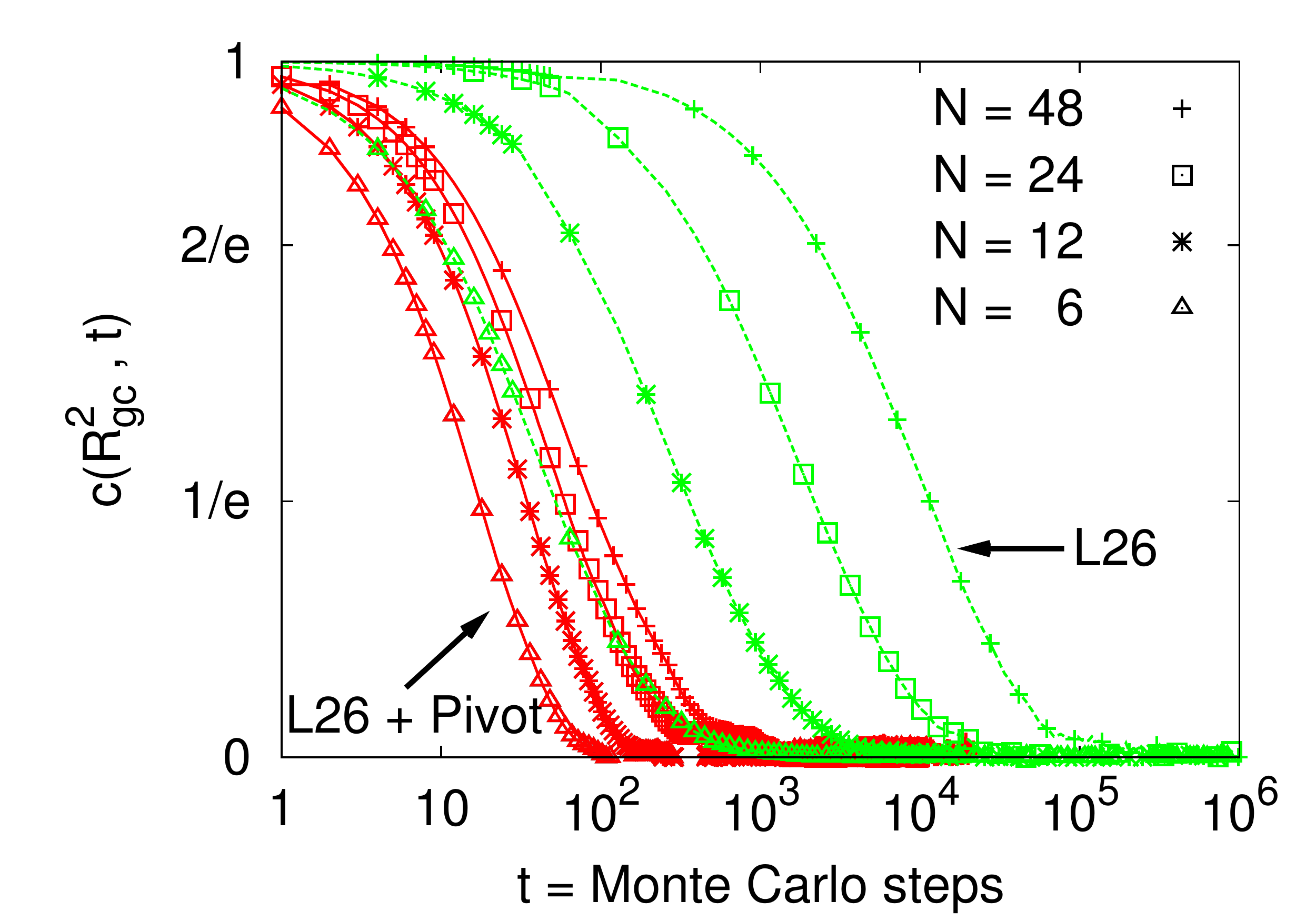}\hspace{0.6cm}
(b)\includegraphics[scale=0.25,angle=270]{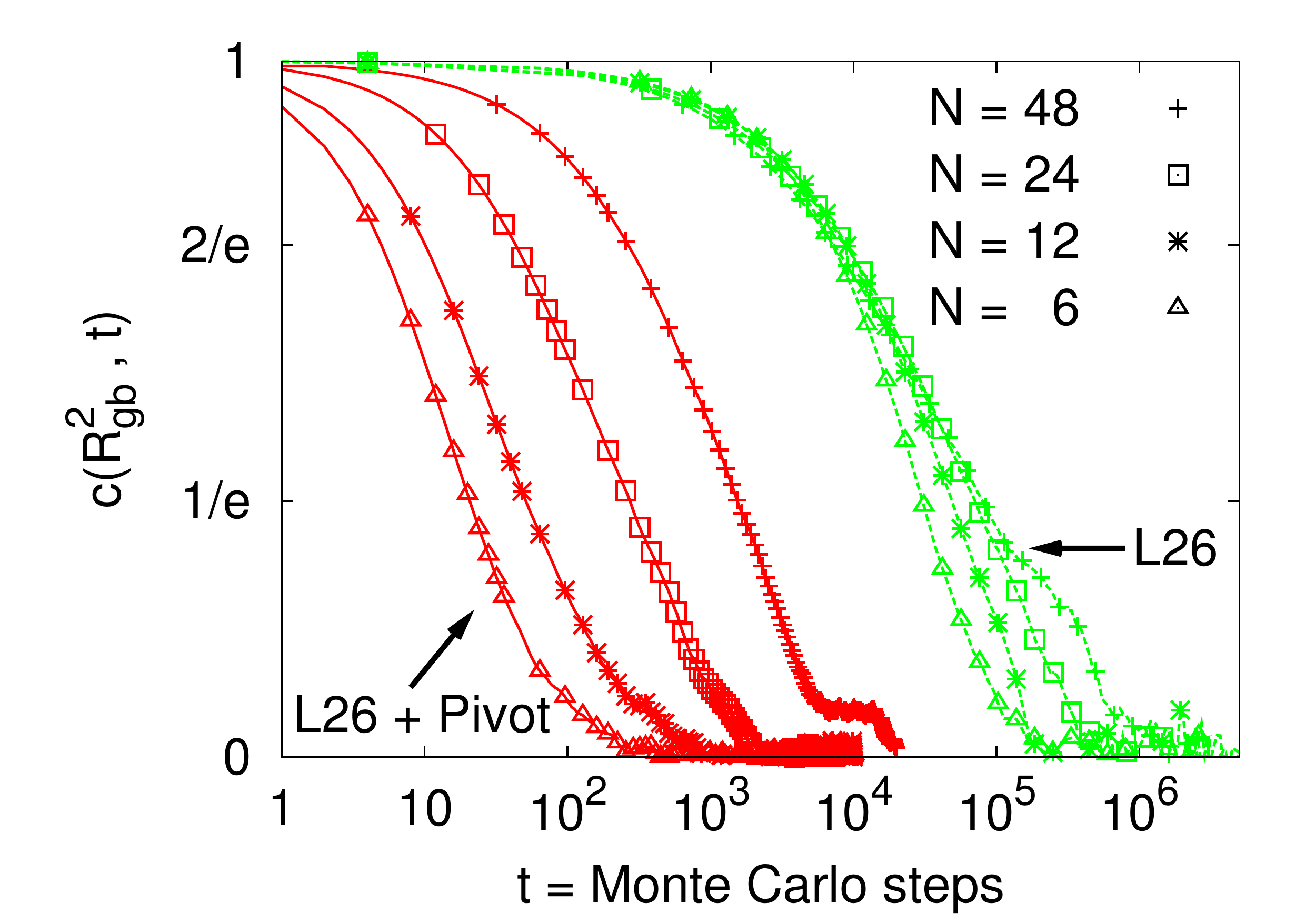}
\caption{Autocorrelation functions of the mean square
gyration radii for the side chains $c(R^2_{gc},t)$ (taking the
average of all side chains at $t$) (a),
and for the backbone $c(R^2_{gb},t)$ (b), plotted vs. the
number of Monte Carlo steps $t$.
Data obtained for bottle-brush polymers with
backbone length $N_b=35$, side chain lengths $N=48$, $24$, $12$, and
$6$, and grafting density $\sigma=1$ when using the
``L26" moves and the ``L26+pivot" moves are shown by the dashed and
solid curves, respectively.}
\label{fig-c-35}
\end{center}
\end{figure}

In our simulations, the lattice size $V=L_x \times L_y \times L_z$
is chosen large enough so that no monomer can interact with itself. 
The $i$th monomer located at the site $\vec{r}_i=(x_i,y_i,z_i)$ is denoted
by $p_i=z_i+y_iL_z+x_i L_y L_z$. No periodic boundary condition is
considered here but a hashing method is used to map the position of 
monomers back into the original box. 
For small bottle-brush polymers it is possible to  
set the lengths of the simple cubic lattice equal
in each dimension, i.e., $L_x=L_y=L_z = 3N_b$ (the maximum length
determined by the initial configuration), 
while for large bottle-brush polymers one encounters difficulties
in the limitation of computer memory. 
Therefore, in order to be able to simulate large bottle-brush polymers,
a new method is introduced which separates the equilibrating process into
several stages. The maximum lattice size which we can use 
on the computer is $2^{28}$. Thus, we first choose 
$L_z=3N_b$, $L_y=L_x$, and $V=L_zL_yL_x \le 2^{28}$. As shown in
Fig.~\ref{fig-bb515-ini}, the time series of $\Delta_x(t)$,
$\Delta_y(t)$, and $\Delta_z(t)$ as well as configurations in the 
intermediate state are stored. After $t_f$ MC steps, we reset
the size of the lattice by decreasing $L_z$ to $L_z=\Delta_z(t_f)$ but 
increasing $L_y$ and $L_x$ to 
$L_y=L_x= {\rm Integer}\{(2^{28}/\Delta_z(t_f))^{1/2}\} 
\ge {\rm max}(\Delta_y(t_f), \Delta_x(t_f))$. 
Here $t_f$ is the number of MC steps at the current stage,
which can be adopted to the simulation such that the condition
$\Delta_x < L_x$ and $\Delta_y < L_y$ holds.
We repeat the same procedure until $L_x=L_y=L_z$ since the space
occupation of the conformations of bottle-brush polymers must be isotropic.

To speed up the equilibrating process and the time for
generating independent configurations,
in addition to the local ``L26'' move~\cite{Wittmer2007}
also pivot moves~\cite{Sokal1995} are used. Two types of moves are
attempted (see Fig~\ref{fig-pivot}): 
\begin{enumerate}
\item[(i)] a monomer on the backbone is chosen randomly and
the short part of the bottle-brush polymer is transformed by
randomly applying one of the 48 symmetry operations (no change;
rotations by 90$^o$ and 180$^o$; reflections and inversions)
to the adjacent bond.

\item[(ii)] A monomer is chosen randomly from all the side chain
monomers, and the part of the side chain from the selected monomer
to the free end of the side chain is transformed by one of the 48
symmetry operations.
\end{enumerate}

Again, we apply the
``L26 + pivot" algorithm to the bottle-brush polymers with
$N_b=515$, $N=12$, and $\sigma=1$ but start the simulations from
three different initial configurations named by e1, e2, and e3.
From the results shown in Fig.~\ref{fig-bb515-ini}, we see that 
the equilibrium states are reached in less than $10^6$ MC steps.
Here one MC step is a sequence of $N_{\rm tot}$ ``L26" moves, 
$k_{\rm pb}$ pivot moves of the backbone
and $k_{\rm pc}$ pivot moves of side chains; $k_{\rm pb}$ is chosen
such that the acceptance ratio is about $40\%$ or even larger, 
while $k_{\rm pc}=n_c/4$. 
It takes about 1.25 hours CPU time on an Intel 2.80 GHz PC to
reach the equilibrium state (after $10^6$ MC steps are performed) 
for one single simulation of bottle-brush polymers ($N_b=515$, $N=12$,
and $\sigma=1$) with $N_{\rm tot}=6671$, $k_{\rm pb}=40$, 
and $k_{\rm pc}=128$.

Another important point one has to be aware of is that
the monomers on the backbone, labelled as 
$0$, $1$, $\ldots$, $N_b-1$, which can be selected  
as a pivot point $N_b^p$ for applying the pivot move are also 
limited due to the
size of the lattice we set up before reaching the equilibrium state.
However, once the system is in equilibrium one has to allow 
all possible moves ($1 \le N_b^p \le (N_b-2)$) with 
the ``L26+pivot" algorithm. For the case shown in
Fig.~\ref{fig-bb515-ini}, the equilibrating process is divided into four 
stages: 

\begin{figure}[htb]
\begin{center}
\vspace{1cm}
(a)\includegraphics[scale=0.25,angle=270]{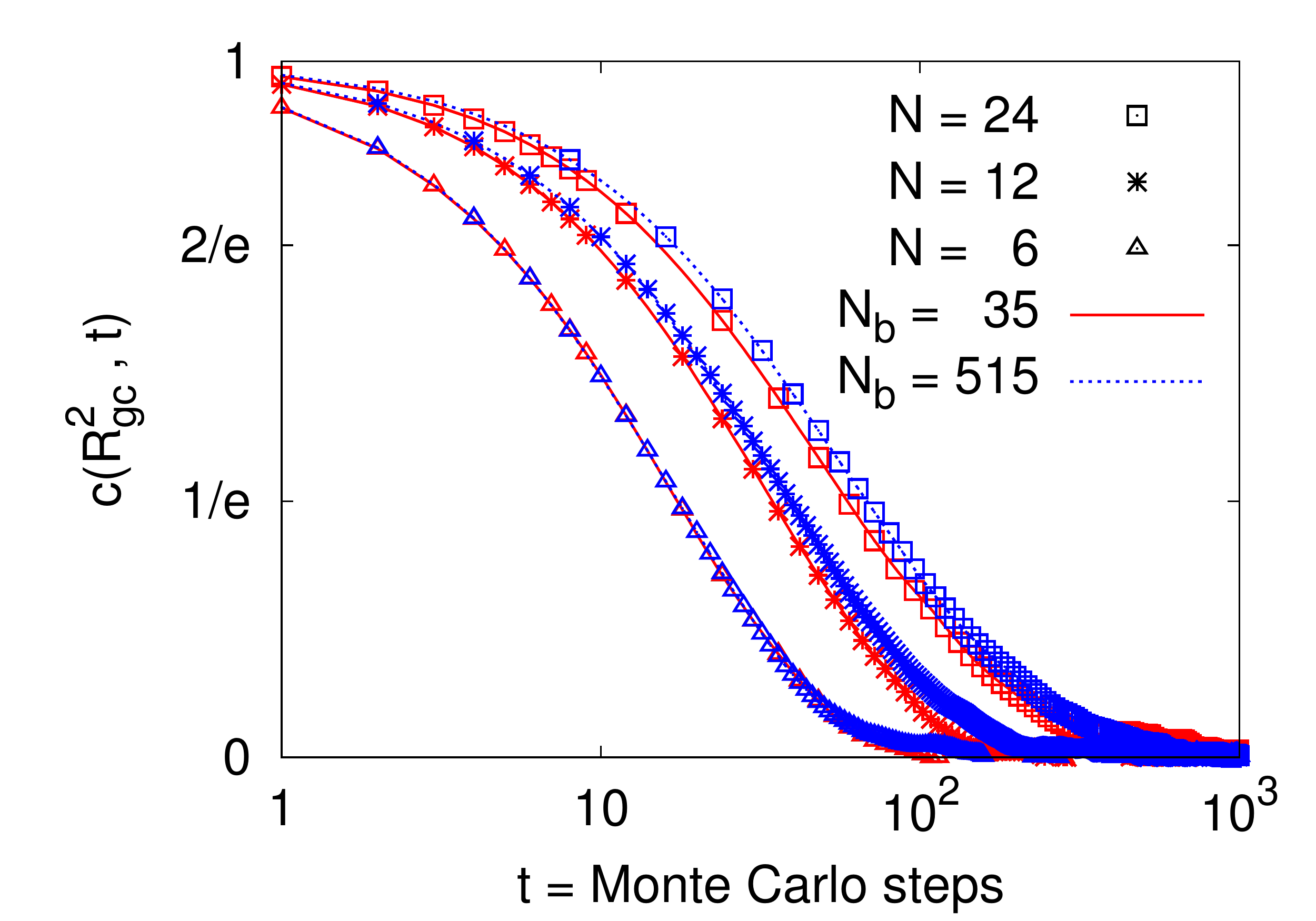}\hspace{0.6cm}
(b)\includegraphics[scale=0.25,angle=270]{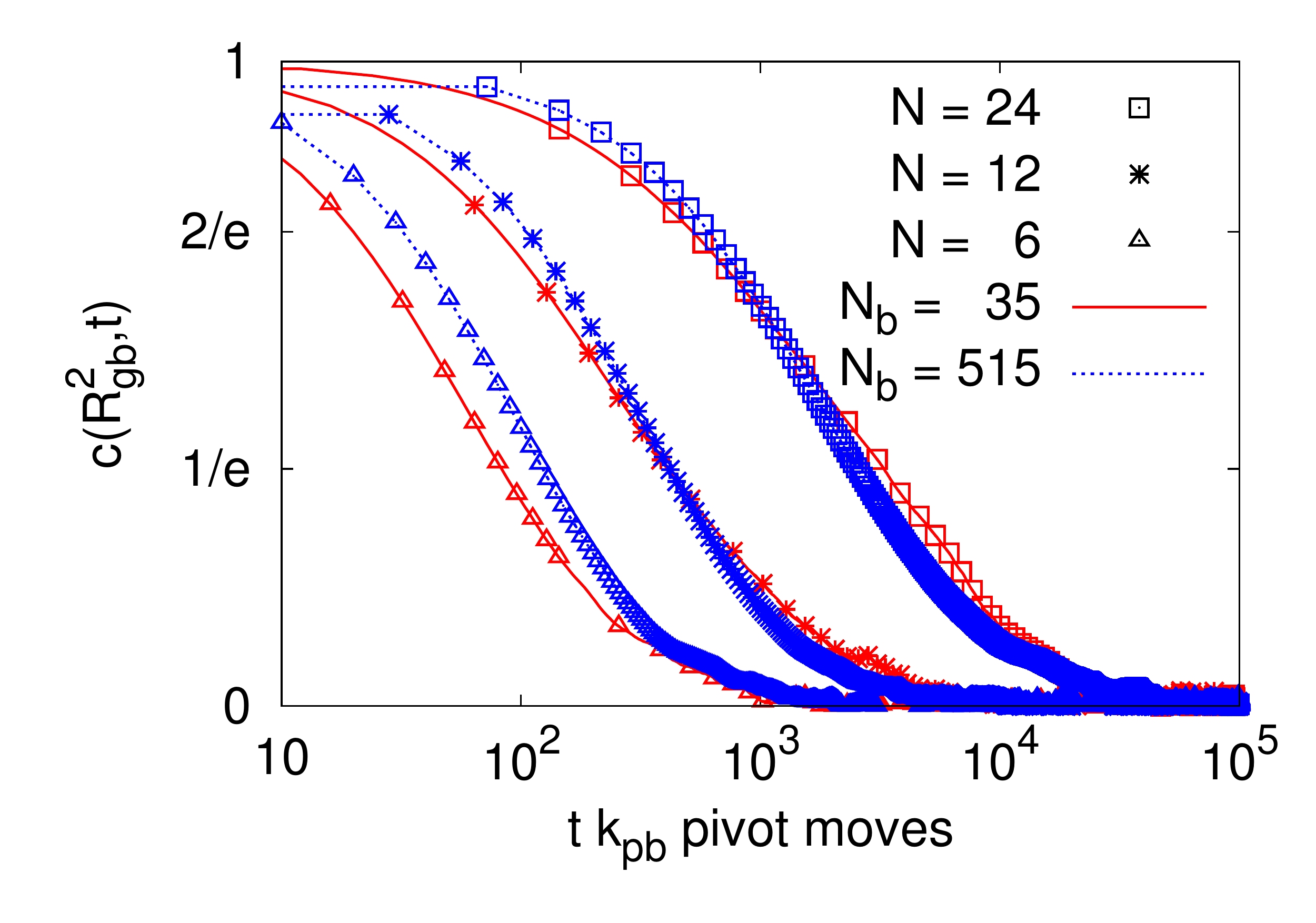}
\caption{(a) Autocorrelation function of the mean square
gyration radii for the side chains $c(R^2_{gc},t)$ as a function of Monte
Carlo time $t$ (a).
Both for $N_b=35$ and $N_b=515$, one pivot move per side chain
is tried on average every $4$ MC steps. (b) Autocorrelation function
for the backbone $c(R^2_{gb},t)$ plotted vs. the total number of pivot moves
done, comparing the relaxation for $N_b=35$ and for $N_b=515$.}
\label{fig-c-515}
\end{center}
\end{figure}

\begin{description}
\item[stage 1:] $1 \le N_b^p \le 128$, 
$L_z=1545$, $L_y=L_x = 415$, $t_f=262144$ MC steps
\item[stage 2:] $1 \le N_b^p \le 256$, $L_z=1201$, 
$L_y=L_x = 473$, $t_f=262144$ MC steps
\item[stage 3:] $1 \le N_b^p \le 513$, $L_z=851$, 
$L_y=L_x = 561$,
$t_f=262144$ MC steps
\item[stage 4:] $1 \le N_b^p \le 513$, $L_z=L_y=L_x = 645$,
$t_f=1310720$ MC steps
\end{description}
As a caveat, the number of stages needed for the system reaching the
equilibrium is empirical. It varies according to the size of 
the backbone length and fluctuations from one equilibration path
to the other equilibration path.
When one simulates an end-grafted bottle-brush polymer adsorbed to an 
impenetrable flat surface, it also varies depending on the attractive
interactions between the monomers and the surface.
Due to the fluctuation of the conformations of bottle-brush polymers, the
time series of the square gyration radius of the backbone,
$R_{gb}^2(t)$, and the space 
occupations in the Cartesian coordinations
$(\Delta_x(t), \Delta_y(t), \Delta_z(t))$ of the bottle-brush polymers
(similar to the functions shown in Fig.~\ref{fig-bb515-ini})
are required in order to determine the lengths of the simulation box
for the next stage and to check whether the equilibrating process is
finished or not. 

The efficiency of the algorithm and the number of
MC steps needed for getting an independent configuration for the
measurement were determined by the autocorrelation function $c(A,t)$,
\begin{equation}
    c(A,t)=\frac{<A(t_0)A(t_0+t)>-<A(t_0)><A(t_0+t)>}
{<A(t_0)^2>-<A(t_0)>^2} \;,
\end{equation}
where $A$ is an observable.
Results of $c(A,t)$ for the mean square gyration radius of the backbone,
$A=R^2_{gb}$, and of the side chains, $A=R^2_{gc}$
(taking the average of all side chains at the same Monte Carlo time $t$)  
plotted against
the number of MC steps $t$ are shown in 
Fig.~\ref{fig-c-35} for $N_b=35$. 
We see that the ``L26+pivot" algorithm is two orders of magnitude faster than
the ``L26" algorithm for the four cases of bottle-brush
polymers (backbone length $N_b=35$, grafting density $\sigma=1$, and 
side chain lengths $N=48$, $24$, $12$, and $6$) we chose. Also, increasing the
side chain length from $N=6$ to $N=48$, the autocorrelation time for the side
chain structural relaxation increases by more than two orders of magnitude for
the ``L26'' case and by less than one order of magnitude for the ``L26+pivot''
case. When we increase the backbone length $N_b$ to $N_b=515$ and adjust the
number of pivot moves for the side chains tried from $k_{\rm pc}=8$
to $k_{\rm pc}=128$ in each Monte Carlo step such
that on average about $1/4$ of the side chains is tried for a pivot move in
each MC step, we can see in part a) of Fig.~\ref{fig-c-515} that the decay of
the autocorrelation function for the side chain structure occurs on the same
time scale, i.e., it is only the average number of pivot moves per side chain
which determines the autocorrelation time. The same is true for the backbone
as we can see in part b) of that figure. Here we plot the autocorrelation
function explicitly against the total number of pivot moves tried, 
$t\, k_{\rm pb}$, and it is obvious that the structural relaxation for 
the much longer backbone
chain occurs on the same time scale as the one for the short backbone chain. A
systematic increase of the relaxation time for both, side chains and backbone,
as a function of side chain length remains, however. 
The stage wise decomposition of the equilibrating process,
with the appropriate adjustment of the simulation volume,
as described above, is a crucial step of our procedures,
and a novel ingredient, not used in other contexts, and
was a prerequisite for successful simulations.

\begin{figure}[htb]
\begin{center}
(a)\includegraphics[scale=0.25,angle=270]{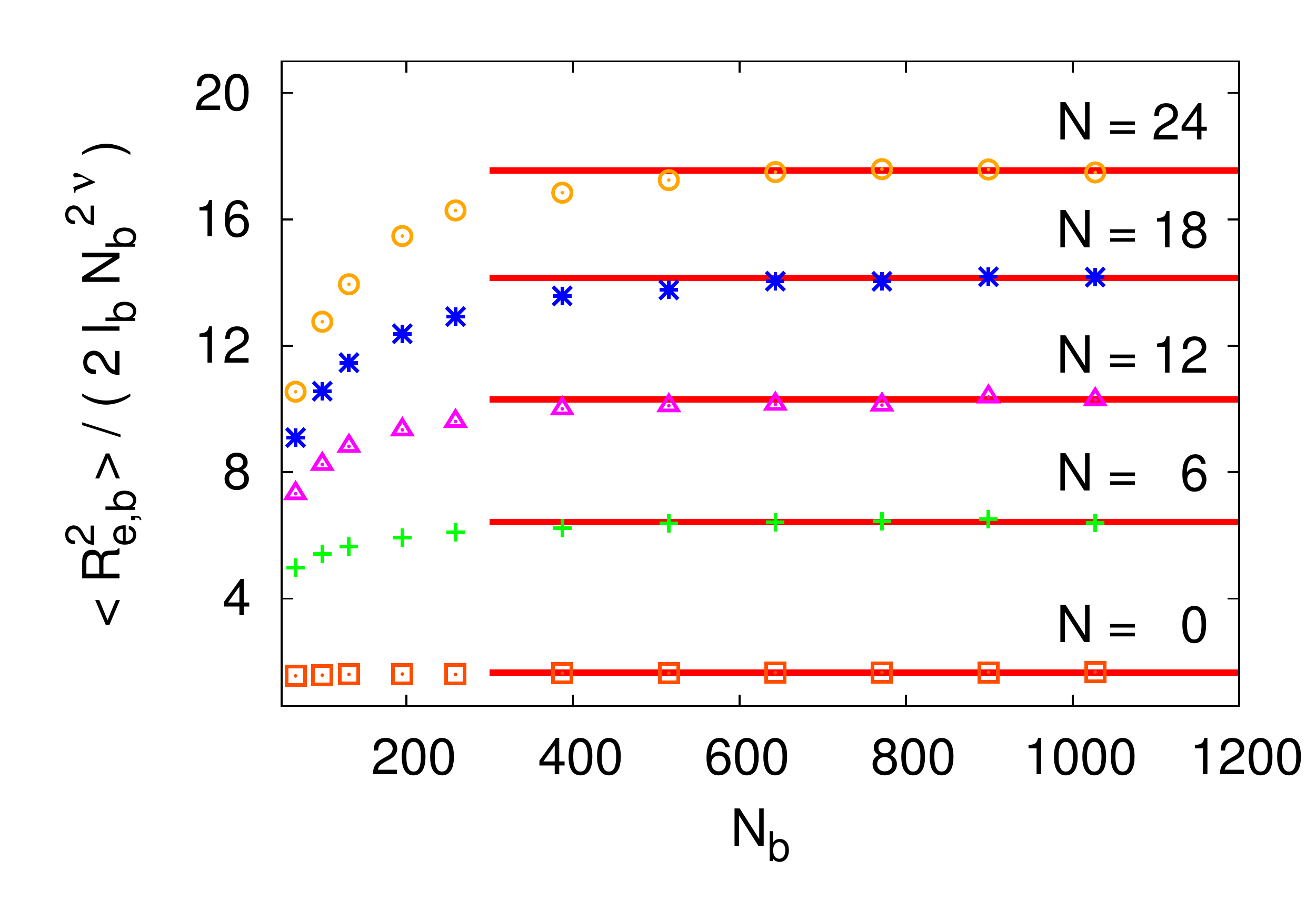}\hspace{0.6cm}
(b)\includegraphics[scale=0.25,angle=270]{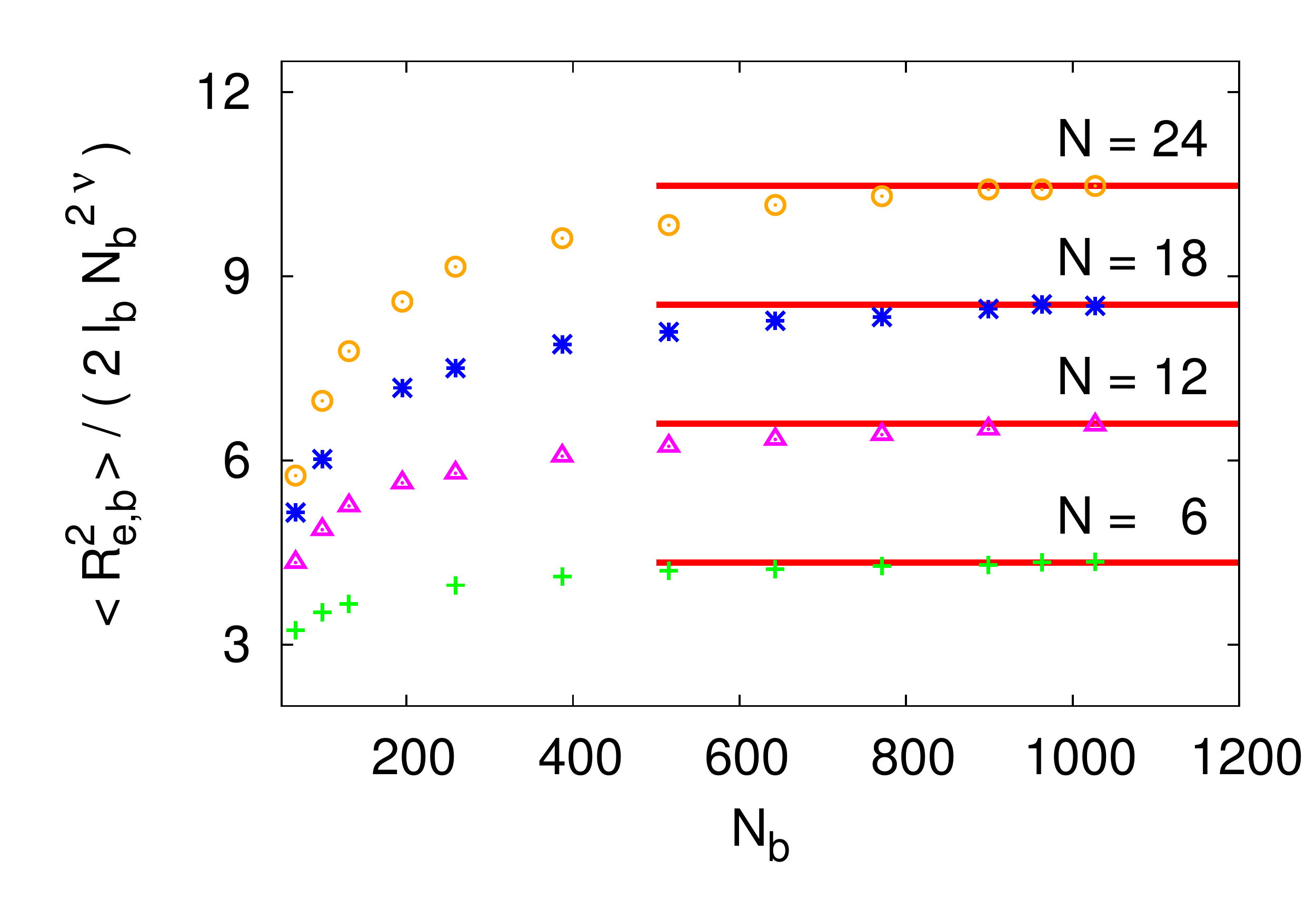}
\caption{\label{fig-Ree} Rescaled mean square
end-to-end distance $\langle R_{eb}^2 \rangle /(2\ell_bN_b^{2\nu})$
($\nu=0.588$) plotted against backbone length $N_b$ for bottle-brush
polymers with grafting density $\sigma = 1$ (a) and $\sigma = 1/2$
(b) Various values of side chain length $N$ are shown (N = 0 means
that no side chains are grafted at all). Horizontal
straight lines indicate the estimates for the persistence length, $\ell_{p,R}$.}
\end{center}
\end{figure}

\begin{figure}[htb]
\begin{center}
\vspace{1cm}
\includegraphics[scale=0.22,angle=0]{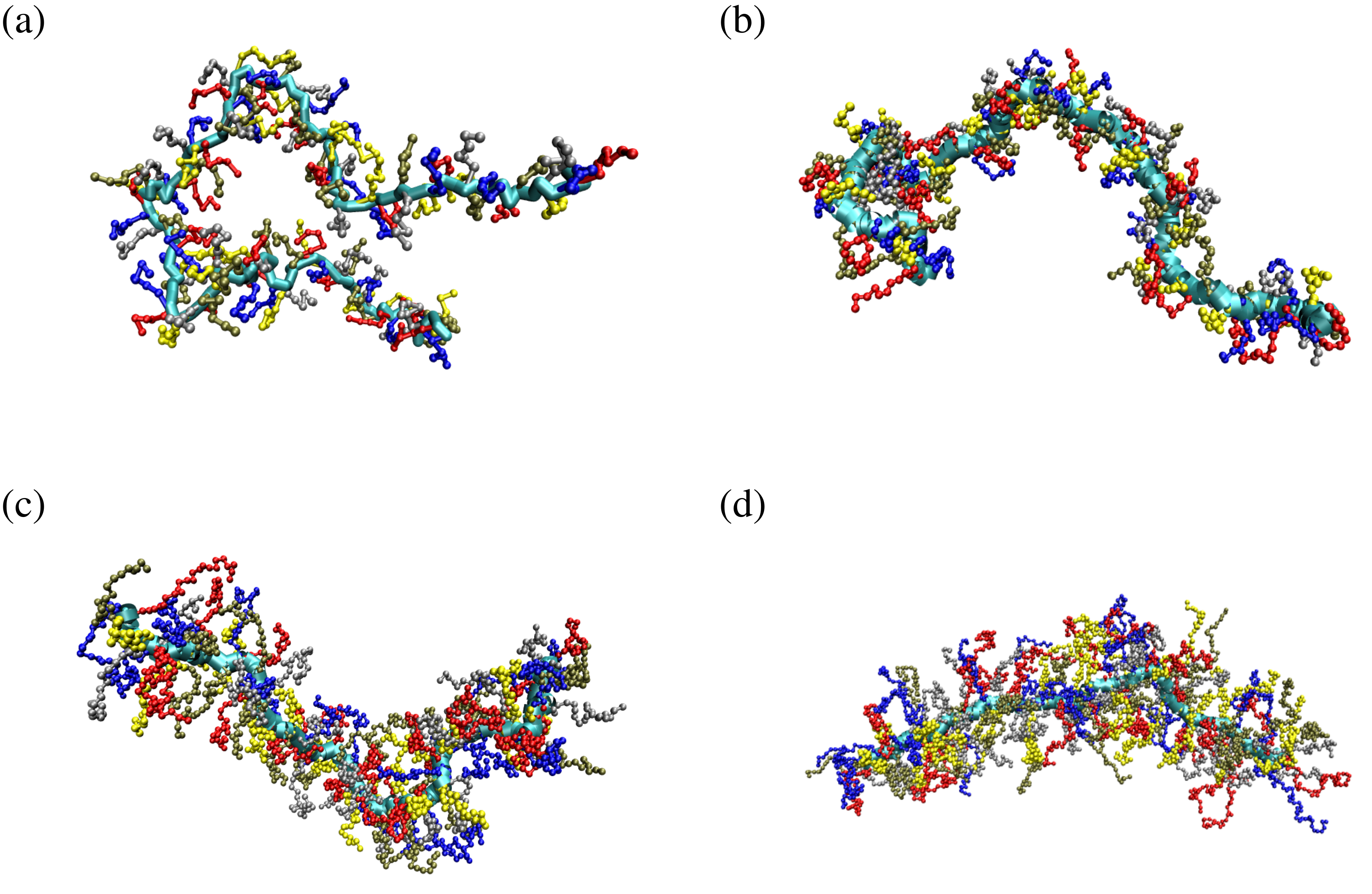}
\caption{Snapshots of the conformations of bottle-brush polymers
with $N_b=131$ monomers on the backbone and with side chain
lengths (a) $N=6$, (b) $N=12$, (c) $N=24$, and (d) $N=48$.}
\label{fig-snapshot}
\end{center}
\end{figure}

\section{Results}
   In order to test our program, we first simulated linear polymer
chains of $N_b$ monomers under good solvent conditions, 
i.e., $N=0$. According to the scaling law of the mean square
end-to-end distance, $R_{eb}^2$, 
one should expect that the curve of 
$ \langle R_{eb}^2 \rangle / N_b^{2 \nu}$ 
becomes horizontal as $N_b \rightarrow \infty$. Here $\nu=0.588$ is 
the Flory exponent for the 3D SAW. 
This is indeed seen in Fig.~\ref{fig-Ree}. 
Using the ``L26+pivot" algorithm, we can simulate bottle-brush
polymers with a number of monomers on the backbone up
to $N_b=1027$, side chain length up to $N=24$ for $N_b>259$,
and side chain length up to $N=48$ for $N_b \le 259$.
The results shown in Fig.~\ref{fig-Ree} are the average of $10^5$-$10^6$
independent configurations. The error bars are given by the
standard deviations of the average, which are smaller than the size
of symbols.
Increasing 
the side chain length $N$, but keeping the grafting density fixed
to $\sigma=1$ or $\sigma=1/2$, one observes an increase in 
$\langle R_{eb}^2 \rangle / N_b^{2 \nu}$ for a fixed value
of $N_b$, which shows that the stretching of the backbone is induced
by the chain length $N$ as well as the grafting density $\sigma$, 
whereas one observes that the backbone 
of the bottle-brush polymers behaves like a SAW
with increasing $N_b$ but fixed $N$. 
In Ref.~\cite{Hsu2010a,Hsu2010b}, it has been
pointed out that the persistence length $\ell_{p,R}$ 
which describes the intrinsic stiffness of bottle-brush polymers
can be determined by the mean square end-to-end distance of the backbone, 
$ R^2_{eb} $, 
\begin{equation}
    \langle R_{eb}^2 \rangle = 2 \ell_{p,R} \ell_b N_b^{2\nu} \;.
\end{equation}
Here $\ell_b \approx 2.7$ is the average bond length of polymer
chains for the bond fluctuation model. 
The snapshots of conformations of 
bottle-brush polymers which contain $131$ monomers on the 
backbone and $N$ monomers on each side chain 
in a good solvent displayed in Fig. \ref{fig-snapshot} also show that
the backbone becomes stiffer as the side chain length $N$ increases
from $6$ to $48$. 

\begin{figure}[htb]
\begin{center}
\vspace{1cm}
(a)\includegraphics[scale=0.25,angle=270]{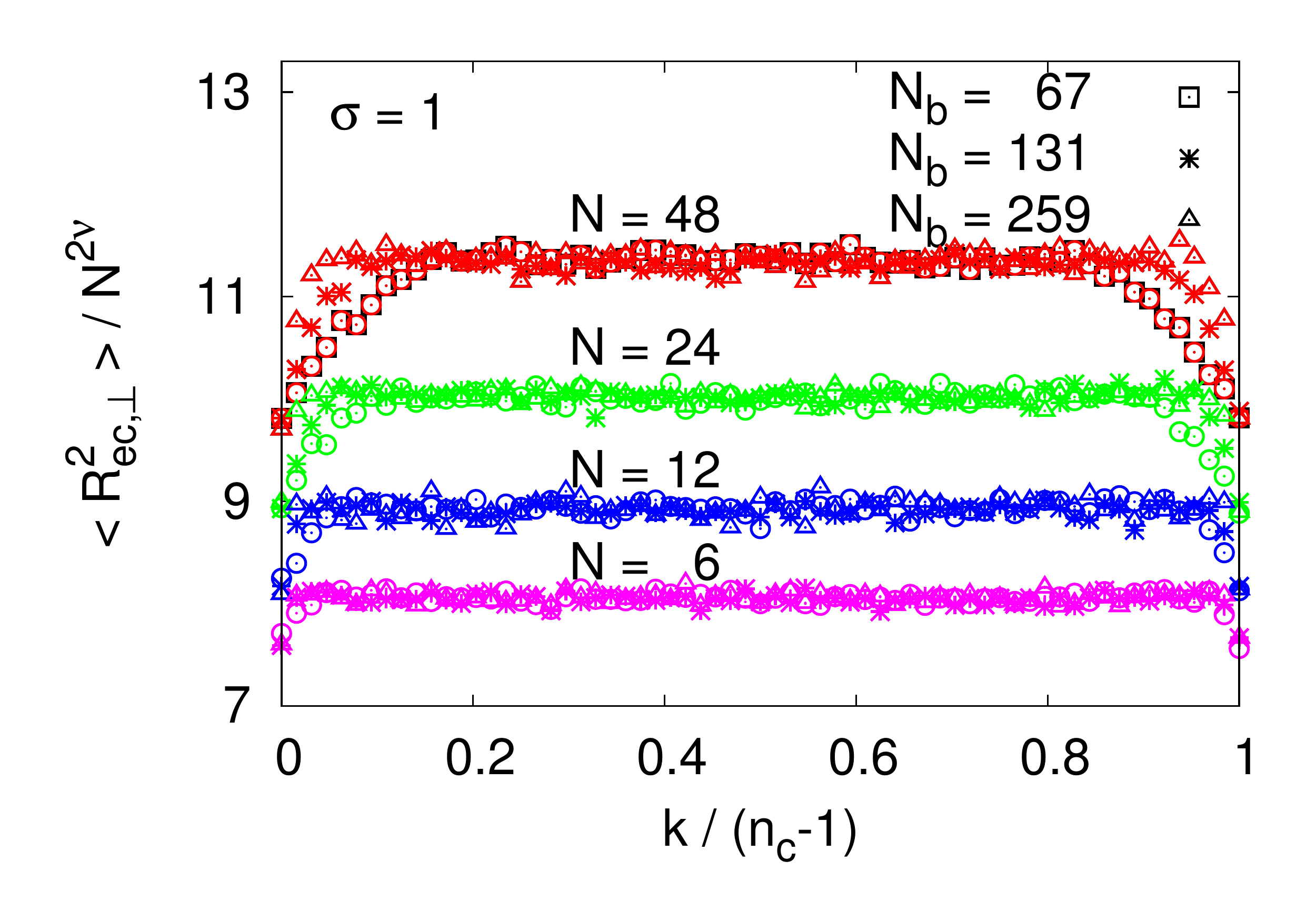}\hspace{0.6cm}
(b)\includegraphics[scale=0.25,angle=270]{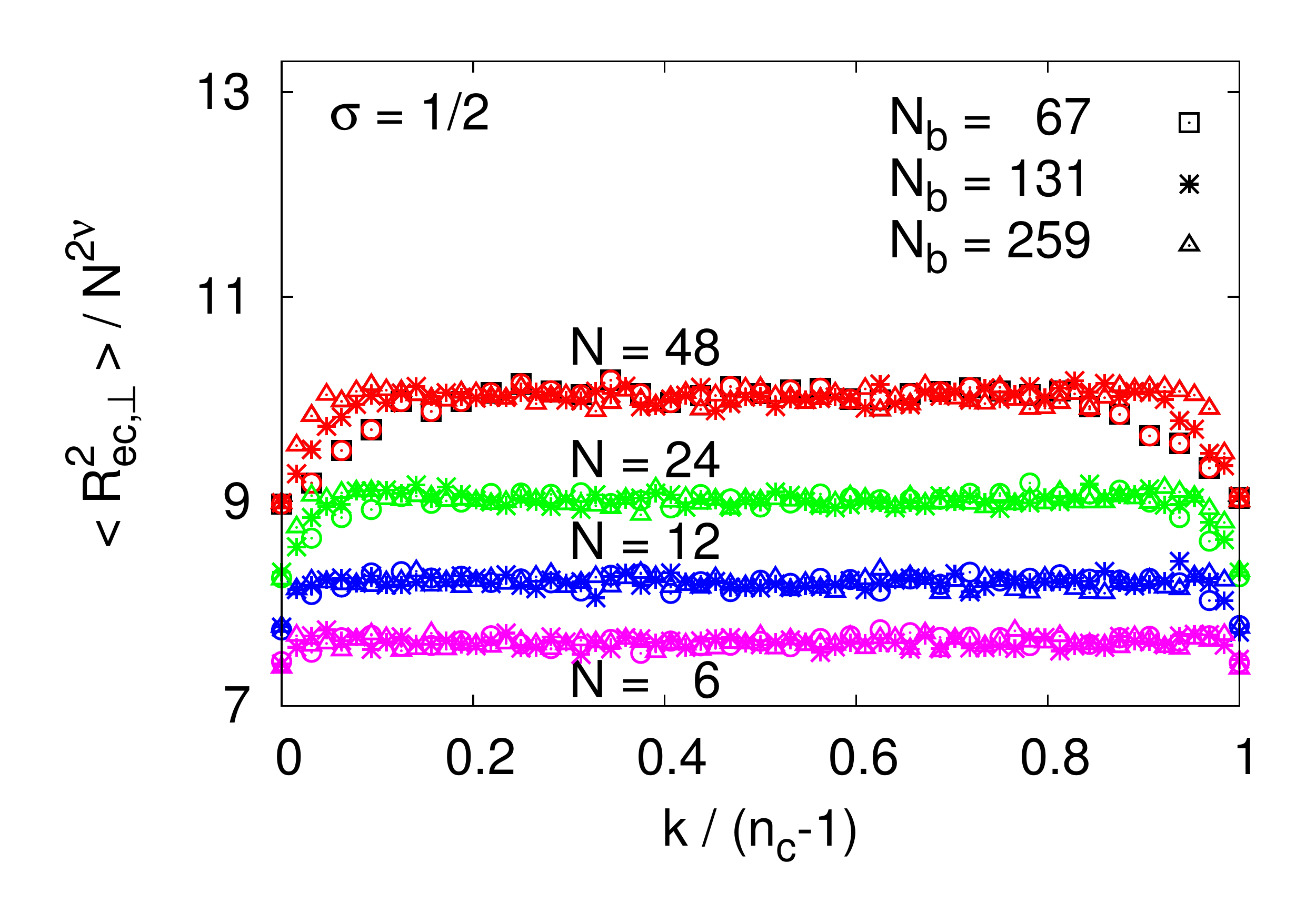}
\caption{Rescaled mean-square end-to-end distance of the side chains in
perpendicular direction $\langle R^2_{ec,\perp} \rangle /N^{2\nu}$
($\nu=0.588$),
plotted vs. the normalized position $k/(n_c-1)$ in the coordinate
system along the backbone (the $n_c$ side chains are labeled by
$k=0$, $1$, $\ldots$, $n_c-1$). Four choices of side chain
length $N=6$, $12$, $24$, and $48$, and three choices of
backbone length $N_b=67$, $131$, and $259$ are included, as indicated.
Case (a) refers to $\sigma=1$, and case (b) to $\sigma=1/2$.}
\label{fig-Recz}
\end{center}
\end{figure}

    Another quantity which can be used to examine whether the system
does reach the equilibrium and the statistics are reliable enough is
the mean square end-to-end distance $\langle R^2_{ec,\perp} \rangle$
(or radius of gyration $\langle R^2_{gc,\perp} \rangle$) of 
the side chains along the backbone
in the direction perpendicular to the backbone. 
For bottle-brush polymers of a flexible backbone,
the perpendicular direction is determined by the vector pointing
from the grafting site to the center of mass of the corresponding 
grafted side chain.
Plotting $\langle R_{ec,\perp}^2 \rangle /N^{2\nu}$ against 
the normalized position 
$k/(n_c-1)$ for $k=0$, $1$, $\ldots$, $n_c-1$ in Fig.~\ref{fig-Recz}, 
a plateau appears with tiny fluctuations for $0.2 < k < 0.8$ 
showing that in the interior of
bottle-brush polymers the side chains behave the same 
for fixed side chain lengths $N$ as expected
after taking the average over sufficiently many independent samples.
The side chains stretch away more from the backbone as the
side chain length $N$ and the grafting density $\sigma$ 
increases. The decreasing behavior of $\langle R^2_{ec,\perp} \rangle$ for
the side chains near the
two ends of the backbone obviously arises from the fact 
that monomers on these side chains
have more freedom to move. 

\begin{figure}[htb]
\begin{center}
\vspace{1cm}
\includegraphics[scale=0.25,angle=270]{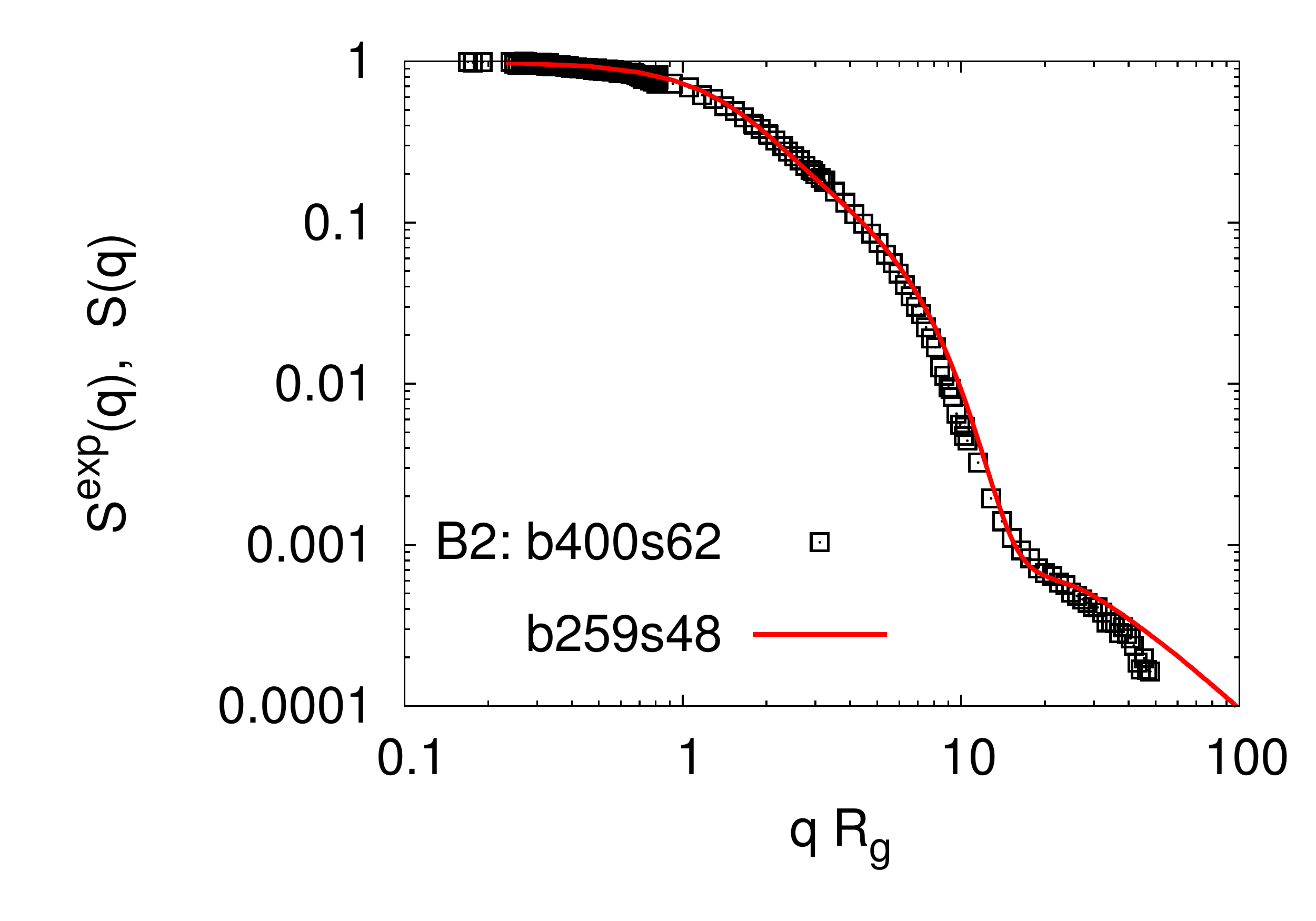}
\caption{Normalized structure factor $S^{\rm exp}(q)$ plotted vs
$qR_g$ for the sample with $N_b^{\rm exp}=400$, $N^{\rm exp}=22$
~\cite{Rathgeber2005}, compared with simulated structure factor $S(q)$
for $N_b=259$, $N=48$~\cite{Hsu2010a}.
$R_g=30.5$ nm obtained from the experimental
data~\cite{Rathgeber2005} and $R_g=115.8$ (lattice spacing) from the
simulation result~\cite{Hsu2010a}. }
\label{fig-exp}
\end{center}
\end{figure}

    Finally, a comparison of the structure factor $S(q)$ 
between the experimental data~\cite{Rathgeber2005} and our 
simulation results~\cite{Hsu2010a} is shown
in Fig.~\ref{fig-exp}. The normalized structure factor $S(q)$ is defined by
\begin{equation}\label{eq1}
S(q) = \frac {1}{N_{\rm{tot}}^2} \sum \limits
_{i=1}^{N_{\rm{tot}}}
\sum\limits_{j=1}^{N_{\rm{tot}}} \langle
c(\vec{r}_i)c(\vec{r}_j)\rangle \frac {\sin (q|\vec{r}_i-
\vec{r}_j|)}{q|\vec{r}_i-\vec{r}_j|},
\end{equation}
where $c(\vec{r}_i)$ is an occupation variable, $c(\vec{r}_i)=1$ if
the site $\vec{r}_i$ is occupied by a bead, and zero otherwise. Note
that an angular average over the direction of the scattering
vector $\vec{q}$ has been performed, and the sums run over all
monomers (all side chains and the backbone).
Adjusting only the number of monomers on the backbone, $N_b$, 
and the number of monomers
on each side chain, $N$, the results for the bottle-brush
polymers of $N_b=259$, $N=48$, and $\sigma=1$ are mapped to the data for the
experimental sample of $N_b^{\rm exp}=400$, $N^{\rm exp}=62$ and 
$\sigma^{\rm exp} \approx 1$
when the momentum  $q$ is scaled by $R_g$. Here 
$R_g$ is the gyration radius of the whole 
bottle-brush polymer. In this case, $R_g=115.8$ (lattice spacings)
is obtained from our MC simulations and $R_g=30.5$ nm is obtained from
the experimental data. One can immediately translate 
$1$ nm $\approx 3.79$ lattice spacing.
Clearly, translating the length units from the large-scale 
structure (small $q$ behavior) and choosing the length of backbone
and side chains correctly, one obtains a faithful description of the 
experimental scattering function over the whole 
$q$-range studied. 
 
\section{Conclusion}
    We have presented extensive MC simulations for
bottle-brush polymers under good solvent conditions using
the bond fluctuation model with a newly developed efficient MC algorithm 
combining the ``L26" moves, the pivot moves, and an adjustable 
simulation lattice box {which changes its shape from a very elongated 
parallelepiped to a cube as the equilibration of the bottle-brush polymer
conformation from an initial stretched backbone to a coiled 
conformation proceeds.}
Using this fast algorithm to generate a sufficiently large number of 
independent configurations in equilibrium, 
we are able to obtain high accuracy
estimates of the related characteristic length scales for describing
the conformational properties of bottle-brush polymers, such as
the end-to-end distance of a side chain and the backbone, 
the persistence length of
the backbone, the effective cross-sectional radius of the whole
bottle-brush polymers etc.
From the computer simulations, 
the scattering intensity contributed by any part of the bottle-brush
polymers are calculated directly. Therefore, we are also able to 
compare our simulation result to the experimental data directly and
test those models used in the analysis of experimental data.

   Recently, this algorithm has also been employed successfully 
to study the conformational change of bottle-brush polymers as they
are absorbed on a flat solid surface by varying the attractive 
interaction between the monomers and the surface~\cite{Hsu2010c}.
In our future work, it will be interesting to see how far
this algorithm can be applied for studying bottle-brush polymers
under poorer solvent conditions.

\section{Acknowledgments}
H.-P. H. received funding from the
Deutsche Forschungsgemeinschaft (DFG), grant No SFB 625/A3. 
We are grateful for extensive grants of computer time at the JUROPA
under the project No HMZ03 and
SOFTCOMP computers at the J\"ulich Supercomputing Centre (JSC).
H.-P. H. thanks J. Baschnagel, H. Meyer, and 
J. P. Wittmer for stimulating discussions during her visit
at the Institut Charles Sadron, Strasbourg, France.
We also thank K. Binder for very useful discussions and for 
critically reading the manuscript.




\bibliographystyle{elsarticle-num}
\bibliography{<your-bib-database>}



\end{document}